\documentstyle[12pt]{article}
\begin{document}
\sloppy
\begin{center}
{\large \bf Quantisation of $O(N)$ invariant nonlinear\\ sigma
model in the Batalin-Tyutin formalism}
\\[1cm]
 N. Banerjee\\ Saha Institute of Nuclear Physics\\1/AF, Bidhannagar, Calcutta
700064\\
 India\\[0.5cm]
 Subir Ghosh\\Gobardanga Hindu college\\North 24-Parganas, West Bengal,
India\\[0.5cm]
                  and\\
 R. Banerjee\\S. N. Bose National Center for Basic Sciences\\Sector I, DB 17,
Bidhannagar\\
Calcutta- 700064, India.\\[1cm]
\end{center}
\begin{abstract}
We  quantise  the  $O(N)$  nonlinear  sigma  model using the Batalin Tyutin
(BT)
approach of converting  a  second  class  system  into  first  class.  It  is a
{\it  nontrivial}  application  of  the BT method since the quantisation of
this
model by  the  conventional  Dirac  procedure  suffers  from  operator ordering
ambiguities.  The  first  class  constraints,  the BRST Hamiltonian and the
BRST
charge are explicitly  computed.  The  partition  function  is  constructed and
evaluated in the unitary gauge and a multiplier (ghost) dependent gauge.
\end{abstract}
\newpage
\section{{\bf Introduction}}

Over  the  last  few  years  a  method  of generalised canonical quantisation
of
constrained  dynamical  systems  has  been  developed   by   Fradkin   and
collaborators  [1,  2] as an alternative to the pioneering formulation of Dirac
[3].
This method [1,2], which has been reasonably well established for  systems with
first class constraints only, has been very recently extended to include
systems
with second class constraints [4,5]. We shall henceforth  refer  to  this later
method as the Batalin-Fradkin (BF) [4] and Batalin-Tyutin (BT) [5] schemes. The
basic idea
of this method is to convert  the  second  class  system  into  first  class by
extending  the  phase  space  and  then to use  the familiar machinary valid
for
first class systems [1,2]. While the BT [5] method remains unexplored (as far
as
quantisation of specific models is concerned),
Some applications of the BF [4] formalism  have  been
reported recently [6--9]. It is noteworthy, however, that these applications
are
confined  to  examples  like  the  chiral gauge theories [6,7], the chiral
boson
theory [8], the massive Maxwell [7] and the massive Yang-Mills [9] theories. In
all  these  models  the  Dirac  brackets  among the canonical variables are
very
simple ({\it i.e.} there are no operator ordering problems) and the
quantisation
can  be,  and indeed has earlier been [10,11], just carried out by the
classical
method of Dirac [3]. Such examples are, therefore, pedagogic  exercises  and do
not  reveal  the  complete  power  or flexibility of either the BF [4] or BT
[5]
approaches. Furthermore, the quantisation  presented  in   ref.  [7]  is  not a
systematic application of the BF or BT method [4,5].

The  motivation  of this paper is to provide a non-trivial application of the
BT
procedure [5]. We shall  consider  the  quantisation  of  the  $O(N)$ invariant
nonlinear  sigma  model. This model, which is a second class system, is known
to
have (quadratic) field dependent Dirac brackets among the  canonical variables.
Consequently  quantisation  by  Dirac's  [3]  procedure is riddled with
operator
ordering ambiguities. The conventional approach is to work in the configuration
space  functional  integral  formalism  [11].  In  this  paper  we  show that
an
ambiguity free operator quantisation of the model can be performed by using the
BT  method  [5]  of  converting  the  second  class system into first class.
The
involutive ({\it i.e.} first class) Hamiltonian contains an infinite  number of
terms,  although  the number of additional  (unphysical)  fields  introduced to
extend the phase space is finite. A remarkable series of cancellations allows
us
to express this  infinite  series  as  a  closed  (exponential)  form. Operator
ordering  problems  never  arise since we always work in the canonical
formalism.
The phase space partition function is next constructed and explicitly evaluated
for two different choices of gauge. In one case (the unitary gauge) the
original
(second class) theory is reproduced. In the other case (ghost dependent gauge)
a
non-trivial structure, which cannot be obtained by conventional [10] phase
space
approach, is obtained.
\vspace{1cm}
\section{\bf Quantisation}

The $O(N)$ nonlinear sigma model consists of a multiplet of $N$ real scalar
fields
$n^{a}$, $a~=~1,\ldots , N$ whose dynamics is governed by the Lagrangian,
\begin{equation}
{\cal L}~=~\frac{1}{4} (\partial_{\mu}n^{a})(\partial^{\mu}n^{a})
\end{equation}
subjected to a primary constraint,
\begin{equation}
T_{1}~=~n^{a}n^{a} - 1 = n^{2} -1 \approx 0.
\end{equation}
The canonical Hamiltonian, obtained by a formal Legendre transform from (1),
is,
\begin{equation}
H_{c} = \Pi_{a}^{2} - \frac{1}{4} \partial_{i}n^{a}\partial^{i}n^{a}
\end{equation}
where $\Pi_{a}$ is canonical momenta,
\begin{equation}
\Pi_{a}   =   \frac{\partial   {\cal  L}}{\partial  \dot{n}^{a}}  = \frac{1}{2}
\dot{n}^{a}.
\end{equation}
Secondary constraints, if present, are found by time conserving $T_{1}$ (2)
with
the total Hamiltonian [3],
\begin{equation}
H_{T} = \int dx \left[H_{c} + \lambda T_{1}\right]
\end{equation}
where  $\lambda$  is  a  Lagrange  multiplier.  Indeed,  there  is  a secondary
constraint $T_{2}$,
\begin{equation}
T_{2} = n^{a}\Pi_{a}\approx 0.
\end{equation}
The  constraints  $T_{1}$  and $T_{2}$ are second  class, satisfying the
Poisson
algebra,
\begin{equation}
\Delta_{\alpha\beta}(x,y)     =      \big\{T_{\alpha}(x)~,~T_{\beta}(y)\big\} =
-2\epsilon_{\alpha\beta}n^{2}\delta(x-y);~~~\alpha,\beta = 1,2
\end{equation}
and   $\epsilon_{\alpha\beta}$   is  the  antisymmetric  tensor  normalised as
$\epsilon_{12} = -\epsilon^{12} = -1$.

Time conserving $T_{2}$, consequently, does not yield a new constraint but
fixes
the multiplier $\lambda$ in (5),
\begin{equation}
\lambda = \Pi^{2} + \frac{1}{4}\partial_{i}n^{a}\partial^{i}n^{a}
\end{equation}
so that the total Hamiltonian (5) becomes,
\begin{equation}
H_{T}  = \int \left[n^{2}\Pi^{2} +
\frac{1}{4}\partial_{i}n^{a}\partial^{i}n^{a}
(n^{2}-2)\right].
\end{equation}
The involutive algebra of the constraints $T_{\alpha}$ with $H_{T}$,
\begin{eqnarray}
\left\{T_{1}(x)~,~H_{T}\right\} &= &4 n^{2} T_{2} \nonumber\\
\left\{T_{2}(x)~,~H_{T}\right\} &= &\frac{1}{4}\partial^{i}\big\{(T_{1}-1)
\partial_{i}
T_{1}\big\} - (\partial_{i}n^{a})(\partial^{i}n^{a})T_{1}
\end{eqnarray}
clearly illustrate the nonlinear features. Indeed the unsystematic approach of
ref.
[7] becomes untenable due to this involved algebra (10).

The  Hamiltonian  (9) with the constraints $T_{\alpha}$ is the starting point
of
our analysis.
The first step is to convert the  second  class  constraints  $T_{\alpha}$ into
first class. In doing this we follow the prescription of ref. [5]. The new
first
class constraints $T^{\prime}_{\alpha}$ are given by,
\begin{equation}
T^{\prime}_{\alpha}  (n^{a},  \Pi_{a},  \phi^{\alpha})   =
\sum_{n=0}^{\infty}
{T^{\prime}_{\alpha}}^{(n)}, ~~~{T^{\prime}_{\alpha}}^{(n)} \sim (\phi)^{n}
\end{equation}
subject to the boundary condition,
\begin{equation}
{T^{\prime}_{\alpha}}^{(0)}   =   T^{\prime}_{\alpha}   (n^{a},  \Pi_{a},  0) =
T_{\alpha}
\end{equation}
and where $\phi^{\alpha}$ are the new dynamical variables in the extended phase
space  $(n^{a},  \Pi_{a})\oplus  (\phi^{\alpha})$ with the basic poisson
algebra [5],
\begin{equation}
\left\{ \phi^{\alpha}(x)~,~\phi^{\beta}(y)\right\} = \omega^{\alpha\beta}(x,y)
\end{equation}
with $\omega$ being an antisymmetric invertible matrix,
\begin{equation}
\omega^{\alpha\beta}(x,y) = -\omega^{\beta\alpha}(y,x).
\end{equation}
After (12), the next term in the series (11) is,
\begin{equation}
{T^{\prime}_{\alpha}}^{(1)} (x) = \int dy ~X_{\alpha\beta}(x,y)
\phi^{\beta}(y)
\end{equation}
where,
\begin{equation}
\int dz~dz^{\prime} \big[ X_{\alpha\mu}(x,z) \omega^{\mu\nu}(z,z^{\prime})
X_{\nu\beta}
(z^{\prime},y)  \big] = - \Delta_{\alpha\beta}(x,y)
\end{equation}
with $\Delta_{\alpha\beta}(x,y)$ defined in (7). The other terms $(n>1)$ in
(11)
are obtained by a recursion relation [5]. As we shall presently see these are
not  needed in our analysis.

A       possible       choice      for      $\omega^{\alpha\beta}(x,y)$ and
$X_{\alpha\beta}(x,y)$ satisfying (14) and (16) is
\begin{eqnarray}
\omega^{\alpha\beta}(x,y) &= &2\epsilon^{\alpha\beta}\delta(x-y) \nonumber\\
X_{\alpha\beta} (x,y) &= &\left(
\begin{array}{lr}
 1 & 0 \\
0 & -n^{2}\\
\end{array}
\right) \delta(x-y).
\end{eqnarray}
There is a  `natural  arbitrariness'  [4,5]  in  this  choice  corresponding to
canonical  transformations  in  the  extended  phase  space. The above choice
is
crucial for simplifying the subsequent algebra. Using (11), (12), (15) and
(17),
the final expressions for the new constraints are,
\begin{eqnarray}
T^{\prime}_{1} &= &T_{1} +  \phi^{1} \nonumber\\
T^{\prime}_{2} &= &T_{2} - n^{2} \phi^{2}
\end{eqnarray}
which are strongly involutive,
\begin{equation}
\left\{ T^{\prime}_{\alpha}(x)~,~T^{\prime}_{\beta}(y)\right\} = 0
\end{equation}
indicating  that  the  terms  in  the  series (11) for $n>1$ are redundant.
This
completes the conversion of the second class constraints $T_{\alpha}$  to first
class  ones  $T^{\prime}_{\alpha}$.  Since the original constraints
$T_{\alpha}$
were in involution with the original Hamiltonian (see (10), the new constraints
$T^{\prime}_{\alpha}$  obviously  violate  this  property.  The  next  step is,
therefore, to compute the new involutive Hamiltonian. Following  ref.  [5], the
general structure of this Hamiltonian can be expressed as a power series,
\begin{equation}
H^{\prime} (n^{a}, \Pi_{a}, \phi^{\alpha}) = \sum_{n=0}^{\infty} H^{\prime
(n)},
{}~~H^{\prime (n)} \sim (\phi)^{n}
\end{equation}
subject to the boundary condition,
\begin{equation}
H^{\prime (0)} = H^{\prime} (n^{a}, \Pi_{a}, 0) = H_{T}
\end{equation}
where the general expression for $H^{\prime (n)}$ is [5]
\begin{equation}
H^{\prime (n+1)} = -\frac{1}{n+1}\int dx dy dz \left[\phi^{\mu}(x)
\omega_{\mu\nu}
(x,y) X^{\nu\rho}(y,z) G_{\rho}^{(n)}(z)\right] ~~(n\geq 0)
\end{equation}
The  matrices  $\omega_{\mu\nu}(x,y)$  and  $X^{\nu\rho}(y,z)$  are  the
inverse
matrices  of  $\omega^{\mu\nu}(x,y)$   and    $X_{\nu\rho}(y,z)$ respectively,
defined  in  (17).  The generating functional $G^{(n)}_{\rho}$ has the
extremely simple form,
\begin{eqnarray}
G_{\rho}^{(0)} &= &\left\{T_{\rho}~,~H_{T}\right\}\nonumber\\
G_{\rho}^{(n)}         &=        &\left\{T_{\rho}^{\prime
(1)}~,~H^{\prime
(n-1)}\right\}_{(n^{a},\Pi_{a})}
+ \left\{T_{\rho}~,~H^{\prime
(n)}\right\}_{(n^{a},\Pi_{a})}~~(n\geq 1)
\end{eqnarray}
the genesis of which is contained in the judicious  choice   (17)  so  that the
series  (11) involves only two terms $T_{\alpha}$ and $T_{\alpha}^{\prime
(1)}$.
Indeed a glance at the general structure for $G_{\rho}^{(n)}$ given in equation
(2.54)   of   [5]   would  convince  the  reader  of  the  remarkable algebraic
simplification  achieved  in  (23).  The  symbol  $\{~,~\}_{(n^{a}, \Pi_{a})}$
appearing  there means that the Poisson bracket has to be evaluated with
respect
to $(n^{a}, \Pi_{a})$.  Using  the  expressions  for  the  original
constraints
$T_{\alpha}$  (2,6)  and  the  Hamiltonian   (9)  as well as (21) to (23), it
is
possible to compute all the terms appearing in the power series  (20). Contrary
to  (11),  the series (20) turns out to be an infinite series. We find,
however,
that a chain of systematic cancellations occurs leading to the result,
\begin{equation}
H^{\prime} = H_{T} - 2\int dx \phi^{2} n^{2}T_{2}  +  \int  dx
\phi^{2}\phi^{2}
(n^{2})^{2} + \sum_{p=1}^{\infty}H^{(p)}
\end{equation}
where,
\begin{eqnarray}
H^{(p)} = \int dx_{1} dx_{2}\ldots dx_{p}[\frac{(-1)^{p}}{p!}& \frac{1}{2}
(\frac{\phi^{1}}{n^{2}}) (x_{1})\{ T_{2}(x_{1})~,~
\frac{1}{2} (\frac{\phi^{1}}{n^{2}}) (x_{2})\{ T_{2}(x_{2})\ldots
\nonumber\\
  &\frac{1}{2}  (\frac{\phi^{1}}{n^{2}}) (x_{p})\{
T_{2}(x_{p})~,~H_{0}\}\}\}_{p-fold}]
\end{eqnarray}
and,
\begin{equation}
H_{0} = H_{T} - \int dx~ n^{2}(x) \Pi^{2}(x)
\end{equation}
is a function of $n^{a}$ fields only. A  convenient  way  to  express  (25) is,
therefore,  to use the functional Schr\"odinger representation $ (\Pi_{a}\to
(-)
\frac{\delta}{\delta n^{a}})$ so that,
\begin{equation}
H^{(p)} =  \frac{1}{p!}\int  dx_{1}\ldots  dx_{p}
\left[\frac{\phi^{1}}{2n^{2}}
\left\{n^{a}\frac{\vec{\delta}_{L}}{\delta n^{a}}\right\}\right]^{p} H_{0}
\end{equation}
where $\vec{\delta_{L}}$ indicates the left derivative. Combining (27) with
(24)
yields the final Hamiltonian,
\begin{equation}
H^{\prime} = \int n^{2}\Pi^{2}  -  2\int \phi^{2}n^{2}T_{2}  +  \int
\phi^{2}\phi^{2} (n^{2})^{2}
+\int \exp \left[\int \frac{\phi^{1}}{2n^{2}}
\left\{n^{a}\frac{\vec{\delta}_{L}}
{\delta n^{a}}\right\}\right]. H_{0}
\end{equation}
which,  by  construction  [5],  is  strongly  involutive  with  the constraints
$T^{\prime}_{\alpha}$:
\begin{equation}
\left\{ H^{\prime}~,~ T^{\prime}_{\alpha}\right\} = 0
\end{equation}
 This  completes  the  conversion  of  the second class system (with
Hamiltonian
$H_{T}$  and  constraints  $T_{\alpha}$)  into  first  class  (with Hamiltonian
$H^{\prime}$  and  constraints  $T^{\prime}_{\alpha}$),  and is one of the
major
accomplishments of our paper.

We now make some comments regarding the construction (28): (i) In passing to
the
quantum   theory  where  Poisson  brackets  are  replaced  by  commutators, the
representation $\frac{\vec{\delta}_{L}}{\delta n^{a}}$ goes over to $(i\hbar)
\frac{\vec{\delta}_{L}}{\delta n^{a}}$, so that (28) is amenable to
perturbation
theory  by  just  expanding  the  exponential;  (ii)  previous  attempts [12]
to
construct an involutive Hamiltonian without extending the phase space led  to a
{\it  nonlocal}  form  whereas  (28)  is  local;  (iii)  since  (28) is
strongly
involutive  (see  (29))  it  implies  [4]  that   the   involutive Hamiltonian
$H^{\prime}$ is identical to the BRST [13] invariant Hamiltonian. Thus,
\begin{equation}
H^{\prime} = H_{BRST}
\end{equation}
The BRST invariance is generated by the BRST charge $Q$ which is given by
\begin{equation}
Q = \int dx \left[C^{\alpha}(x)T^{\prime}_{\alpha}(x) + p_{\alpha}(x)
P^{\alpha}
(x)\right];~~ \alpha = 1,2
\end{equation}
where $(C^{\alpha}~,~\bar{P}_{\beta})$ and $(P^{\alpha}~,~\bar{C}_{\beta})$
form
a  canonical  ghost (antighost) pair      having the opposite Grassman parity
as
$T_{\alpha}$;
\begin{equation}
\left\{ C^{\alpha}(x)~,~\bar{P}_{\beta}(y)\right\} = \left\{ P^{\alpha}(x)~,~
\bar{C}_{\beta}(y)\right\}  =  \delta^{\alpha}_{\hphantom{\alpha}\beta}
\delta (x-y)
\end{equation}
while  $(p_{\alpha}~,~q^{\beta})$  is a canonical multiplier pair with
identical
Grassman parity to $T_{\alpha}$;
\begin{equation}
\left\{ q^{\alpha}(x)~,~{p}_{\beta}(y)\right\}
= \delta^{\alpha}_{\hphantom{\alpha}\beta}   \delta (x-y)
\end{equation}
The fields $\bar{P}_{\alpha},\ \bar{C}_{\alpha},\ q_{\alpha}$ do  not  occur in
(31) but will used later.

The BRST charge generates the following transformations $(\delta_{Q}{\cal O} =
\{
Q~,~{\cal O}\})$ on the canonical variables in the complete extended space,
\begin{eqnarray}
\delta_{Q}  n^{a}  =  -C^{2}n^{a}  ~~&~~\delta_{Q} \Pi_{a} = 2C^{1}n^{a} +
C^{2}
(\Pi_{a} - 2n^{a}\phi^{2})\nonumber\\
\delta_{Q}\phi^{1} = 2C^{2}n^{a} ~~&~~ \delta_{Q}\phi^{2}= 2C^{1}\nonumber\\
\delta_{Q} \bar{P}_{\alpha} = T^{\prime}_{\alpha} ~~&~~  \delta_{Q}C_{\alpha} =
0\nonumber\\
\delta_{Q}\bar{C}_{\alpha}  = p_{\alpha} ~~&~~ \delta_{Q}P_{\alpha} =
0\nonumber \\
\delta_{Q} q_{\alpha} = -P_{\alpha} ~~&~~ \delta_{Q}p_{\alpha} = 0
\end{eqnarray}
under which the  BRST  invariance  of  (30)  can  be  explicitly  verified. The
nilpotency condition  $\delta_{Q}^{2}  =  0$   is  clearly  preserved  in (34).
Finally, the physical space is defined by
\begin{equation}
Q|phys\rangle = 0, ~~~~~~~~|phys\rangle \neq  Q|...\rangle
\end{equation}
This  completes  the  operator  formulation  of  the model. We next consider
the
partition function. The first step is  to  define  the  gauge  fermion operator
$\psi$ given in ref.[4]
\begin{equation}
\psi  = \int dx \left[\bar{P}_{\alpha}q^{\alpha} +
\bar{C}_{\alpha}\chi^{\alpha}
\right] \end{equation}
where $\bar{P}_{\alpha},\ \bar{C}_{\alpha},\ q^{\alpha}$ have  been  defined in
(32,33)  and  $\chi_{\alpha}$ is the hermitean gauge fixing function with
identical
Grassman parity as $T_{\alpha}$ and satisfy,
\begin{equation}
det|\{ \chi_{\alpha}~,~T^{\prime}_{\beta}\}|\neq 0
\end{equation}
The complete unitarising Hamiltonian $H_{U}$ is now defined as,
\begin{equation}
H_{U} = H_{BRST} + \{\psi~,~Q\}
\end{equation}
which is also BRST invariant since the added term is  a  BRST  total derivative
[14]. We  now rename the variables $\phi^{1}$ and $\phi^{2}$ as,
\begin{equation}
\phi^{1}\to 2\phi,~~~~~~~~~\phi^{2}\to \Pi_{\phi}
\end{equation}
so  that $(\Pi_{\phi}~,~\phi)$ can be regarded as a canonically conjugate pair
by
virtue of (13) and (17). The partition function $Z$ may now be written as,
\begin{equation}
Z = \int [{\cal D}\mu] e^{iS}
\end{equation}
where,
\begin{equation}
S   =   \int   \left[   \Pi_{a}    \dot{n}^{a}    +    \Pi_{\phi}\dot{\phi} +
C^{\alpha}\dot{\bar{P}}_{\alpha} + P^{\alpha}\dot{\bar{C}}_{\alpha}  +
p_{\alpha}
\dot{q}^{\alpha} - H_{U}\right]
\end{equation}
and the measure $[{\cal D}\mu]$ includes all the variables occuring in the
action.

Let us now explicitly compute $Z$  in  different  gauges.  First,  consider the
`unitary  gauge' [4,5] where the gauge conditions are just the original set of
second
class constraints,
\begin{equation}
\chi_{\alpha} = T_{\alpha}
\end{equation}
Making  the   change   of   variables   $\chi_{\alpha}\to \chi_{\alpha}/\beta$,
$p_{\alpha}  \to  \beta p_{\alpha}$, $\bar{C}_{\alpha}\to
\beta\bar{C}_{\alpha}$
whose (super) Jacobian is unity [1, 14], and finally taking the limit
$\beta\to
0$ [1,14], we obtain,
\begin{equation}
Z  =  \int  \left[{\cal  D}n^{a} {\cal D}\Pi_{a} {\cal D}\phi {\cal
D}\Pi_{\phi}
\right] \delta (T_{1}) \delta(T_{2}) \delta(\phi)   \delta(\Pi_{\phi})
 det |-2n^{2}| e^{iS}
\end{equation}
with,
\begin{equation}
S = \int\left(\Pi_{a}\dot{n}^{a} + \Pi_{\phi}\dot{\phi} - n^{2}\Pi^{2} +  2
\Pi_{\phi} n^{2} T_{2}
-\Pi_{\phi}^{2}   (n^{2})^{2}\right) +  \exp\int \left\{\frac{\phi}{n^{2}}
n^{a}
\frac{\delta}{\delta n^{a}}\right\} H_{0}.
\end{equation}
Note that due to $\delta (T_{1})$ in (43) the Faddeev-Popov determinant reduces
to  a  constant  which  can  be  absorbed  in  the normalisation. The $\phi$
and
$\Pi_{\phi}$ integrals can be trivially performed.  Finally  $\delta(T_{2})$ is
expressed by its corresponding Fourier transform leading to an action,
\begin{equation}
S = -\Pi^{2} + \Pi_{a} (\dot{n}^{a} - \xi n^{a}) - \frac{1}{4}
\partial_{i}n^{a} \partial^{i}n^{a}
\end{equation}
where  $\xi$  is  the  Fourier variable. The Gaussian integral over $\Pi_{a}$
is
done yielding,
\begin{equation}
Z=  \int  {\cal  D}n^{a}  \delta  (n^{2}  -1)   \exp   \left[i\int \frac{1}{4}
(\partial_{\mu} n^{a})(\partial^{\mu}n^{a})\right]
\end{equation}
where  we  have  absorbed  a trivial Gaussian over $\xi$ into the
normalisation.
Expression (46) is, therefore,  seen  to  reproduce  the  original  theory (1)
subject to the constraint (2).

Finally  we  show  how  nontrivial  consequences  arise  from (40) by choosing
a
multiplier dependent gauge,
\begin{equation}
\chi_{1} = \Pi_{\phi} + p_{1}, ~~~~ \chi_{2} = \phi
\end{equation}
As  usual,  scaling  arguments  [1,14]  once  again   enforce   the constraint
$T^{\prime}_{2}$ and the gauge condition $\chi_{2}$ by delta functions.
The  presence  of  the  multiplier   in $\chi_{1}$ prevents this enforcement
for
$T^{\prime}_{1}$ and $\chi_{1}$. We find,
\begin{equation}
Z  =  \int  \left[{\cal  D}n^{a} {\cal D}\Pi_{a} {\cal D}\phi {\cal
D}\Pi_{\phi}
{\cal D}q^{1}{\cal D}p_{1}\right]
\delta(T^{\prime}_{2}) \delta(\phi)
 det |-n^{2}| e^{iS}
\end{equation}
where,
\begin{eqnarray}
S =\int(\Pi_{a}\dot{n}^{a}~ +~ \Pi_{\phi}\dot{\phi}~+~p_{1}\dot{q}^{1}
- & n^{2}~\Pi^{2}
 +  ~ \Pi_{\phi}^{2}   (n^{2})^{2}  -  \exp\int \left\{\frac{\phi}{n^{2}} n^{a}
\frac{\delta}{\delta n^{a}}\right\} H_{0}\nonumber\\
  & +q^{1}(n^{2} - 1 + 2\phi) + p_{1}(\Pi_{\lambda} + p_{1}))
\end{eqnarray}
The $\phi$ integration is trivially done. A Fourier  transformed representation
for  $\delta(T^{\prime}_{2})$   is  taken. Doing successively the Gaussians
over
$p_{1}$, $\Pi_{a}$, $\Pi_{\phi}$ and the Fourier variable $\xi$ yields,
\begin{equation}
Z = \int {\cal D}n^{a} {\cal D}q^{1} (det|-n^{2}|)^{1/2} e^{iS}
\end{equation}
with,
\begin{eqnarray}
S~=~ \int [q^{1} (n^{2}-1) ~ +~ \frac{1}{4n^{4}}~ \big\{
n^{2}(\dot{n}^{a})^{2}
&+  (4n^{4}-1)~(n^{a}  \dot{n}^{a})^{2}\big\} -
{}~(n^{a}\dot{n}^{a})~\dot{q}^{1}
\nonumber\\
& -\frac{1}{2} ~ \partial_{i} n^{a} \partial^{i} n^{a} (1- \frac{n^{2}}{2})]
\end{eqnarray}
Expression (50)apparently differs drastically from (46) but both are expected
to
yield  identical  S-matrix  elements  by  the Fradkin-Vilkovisky theorem
[1,14].
Indeed the nontrivial structure (50)  illustrates  the  generality  of  the
this
approach [4,5] since it cannot be obtained by conventional [10, 15]
quantisation
methods. Note the explicit presence of the  Faddeev-Popov  determinant  in (50)
which, in the previous case (46), could be absorbed in the normalisation.

\vspace{1cm}
\section{\bf Conclusion}

To  conclude,  we  have shown how the BT method [5] can be exploited to
quantise
the $O(N)$ invariant sigma model. In our  knowledge  this  is  the  {\it first}
nontrivial  (in the sense that the conventional Dirac [3] method is riddled
with
operator ordering problems) application of the generalised canonical approach
[4,5].
Moreover since the present example does not involve the gauge field the
conventional
St\"uckelberg [16] or Wess-Zumino [17] approaches of obtaining a first class
theory
is not straightforward. As is  usual  in
such  an  explicit  analysis, new theoretical insights have been gained. We
have
seen the necessity of making an intelligent choice in (17) which simplifies the
algebra  remarkably  and allows us to identify a canonically conjugate pair
(39)
among the new variables. Moreover we find that, contrasted with [4], the recent
work  [5]  is  better  suited  for  computational  reasons.  This is because,
by
construction, it automatically yields a strongly involutive system which is
BRST
invariant.  There  is  another important aspect which we wish to stress. The
generalised
approach [4,5] does not specify the precise Hamiltonian  ({\it  i.e.} canonical
(3)  or  total  (9)) to take as the starting Hamiltonian. We have found that
the
total Hamiltonian (9) is  the  better  choice  since  it  leads  to  the closed
(exponential) form for the involutive Hamiltonian (28). In fact this
Hamiltonian
is  also  ideal  for  perturbative  computations.
Moreover, in contrast with the earlier work [12], this expression is local.
A   corresponding   analysis
originating  from  the  canonical  Hamiltonian  (3)  leads  to  severe
algebraic
complications. Since our analysis is quite  general  it  could  be  employed to
quantise  other  types of nonlinear sigma models ($CP^{N}$ models, for
instance)
including their supersymmetric generalisations.

\vspace{1cm}
\subsection*{Acknowledgements}
 One of the authors (R. B.) thanks the Alexander  von  Humboldt  foundation for
providing  financial support during his stay at Heidelberg, where a part of
this
work was done.


\newpage
\begin{thebibliography}{99}
\bibitem{1} E. S. Fradkin and G. A. Vilkovisky, Phys. Lett. B55 (1975) 224\\
CERN Report TH-2332 (1977) (unpublished).
\bibitem{2} I. A. Batalin and E. S. Fradkin,
Phys. Lett. B122 (1983) 157\\ ibid B128 (1983) 303.
\bibitem{3}   P.  A.  M.  Dirac,  ``Lectures  on  quantum  mechanics'' (Yeshiva
University Press, New York 1964).
\bibitem{4} I. A. Batalin and E. S. Fradkin, Nucl. Phys. B279 (1987) 514; Phys.
Lett. B180
(1986) 157.
\bibitem{5} I. A. Batalin and I. V. Tyutin, Int. J. Mod Phys. A6 (1991) 3255.
\bibitem{6} M. Moshe and Y. Oz, Phys. Lett. B224 (1989) 145\\
P. P. Srivastava, ibid, 235 (1990) 287.
\bibitem{7} T. Fujiwara, Y. Igarashi and J. Kubo, Nucl. Phys. B341 (1990)
695\\
Y. Kim, S. Kim, W. Kim, Y. Park, K. Kim and Y. Kim, Phys. Rev. D46 (1992) 4574.
\bibitem{8}  J.  Kowalski-Glikman,  Phys.  Lett.  B245  (1990)79\\   S. Ghosh,
communicated to Phys. Rev. D.
\bibitem{9} I. A. Batalin and I. V. Tyutin, Mod. Phys. Lett. A7 (1992) 1931.
\bibitem{10} P. Senjanovic, Ann. Phys. [N.Y.] 100 (1976) 227.
\bibitem{11}  For a review, see E. Abdalla, M. C. B. Abdalla and K. D. Rothe,
``
Non  perturbative  methods  in  2  dimensional  quantum  field  theory'' (World
scientific, 1991).
\bibitem{12} P. Mitra and R. Rajaraman, Ann. Phys. [N.Y.] 203 (1990) 137.
 \bibitem{13}  C.  Becchi,  A.  Rouet and R. Stora, Ann. Phys. [N. Y.] 98
(1976) 287\\
 I. V. Tyutin, Lebedev preprint 39 (1975).
\bibitem{14} M. Henneaux, Phys. Rep. C126 (1985) 1 and  ``Classical foundations
of BRST symmetry'' (Bibliopolis, Naples, 1988).
\bibitem{15} L. D. Faddeev, Theor. Math. Phys. 1 (1970) 1.
\bibitem{16} E. C. G. St\"uckelberg, Helv. Phys. Acta 30 (1957) 209.
\bibitem{17} J. Wess and B. Zumino, Phys. Lett. B37 (1971) 95.
\end{thebibliography}
\end{document}